\documentclass[aps,prl,twocolumn,showpacs]{revtex4}
\usepackage{epsf}

\begin{document}

\title {
Kondo screening in unconventional superconductors: \\
The role of anomalous propagators
}

\author{Lars Fritz and Matthias Vojta}
\affiliation{\mbox{Institut f\"ur Theorie der Kondensierten Materie,
Universit\"at Karlsruhe, 76128 Karlsruhe, Germany}}
\date{July 27, 2005}

\begin{abstract}
The Kondo effect in superconductors is frequently investigated
using the local quasiparticle density of states as sole bath characteristics,
i.e., the presence of anomalous propagators is ignored.
Here we point out that this treatment is {\em exact} for a number of
situations, including point-like impurities in $d$-wave superconductors.
We comment on recent investigations [M. Matsumoto and M. Koga,
J. Phys. Soc. Jpn. {\bf 70}, 2860 (2001) and Phys. Rev. B {\bf 65}, 024508 (2002)]
which reached different conclusions:
while their numerical results are likely correct, their interpretation
in terms of two-channel Kondo physics and an ``orbital effect of Cooper pairs''
is incorrect.
\end{abstract}

\maketitle


The physics of quantum impurity moments in superconductors,
associated with the Kondo effect,
has been subject of numerous investigations in recent years.
Diverse theoretical techniques have been employed to study Kondo or
Anderson models in a superconducting environment
\cite{sakaisc,withoff,cassa,tolya,MVRB,MVMK,koga1,koga2,koga3}.
Most of these studies effectively neglect the presence of superconducting (sc)
fluctuations in the host, i.e., they use the local fermionic density
of states (DOS) as the only input quantity characterizing the environment
of the Kondo impurity.
For $s$-wave superconductors it has been shown \cite{sakaisc}
that this approximation is not justified:
here the properties of a Kondo impurity are different from the ones
of an impurity embedded in a non-superconducting system
with the same DOS.
More precisely, the superconducting bath turns out to be equivalent to
a non-superconducting bath with additional particle-hole (p-h) asymmetry \cite{sakaisc}.
As a result, for a p-h symmetric conduction band
a screened singlet state {\em is} realized at large Kondo coupling,
in contrast to the non-sc case with a hard gap in the
local DOS \cite{hardgap}.
This difference can be understood as caused by the anomalous bath
propagators.

In this Brief Report, we address the role of anomalous propagators in unconventional
superconductors. We argue below that neglecting sc propagators is {\em exact}
in many, potentially experimentally relevant, cases,
e.g., for point-like impurities in $d$-wave and unitary $p$-wave superconductors.
In these situations, the dynamics of the impurity degrees of freedom
can be calculated using the normal bath propagators only.
We also comment on recent papers by Matsumoto and Koga who
argued in favor of a non-trivial ``orbital effect of Cooper pairs''
for point-like impurities in both $p+ip$ and $d+id$ superconductors \cite{koga1,koga2}.
While we believe that their numerical results are correct, we point out that
such an orbital effect does not exist: in their situation the impurity properties
are exclusively determined by the local DOS.

We start from the action of an Anderson impurity in a general interacting host.
\begin{eqnarray}
{\cal S} &=& \frac{1}{\beta\cal{N}}
             \sum_{\omega_n} \sum_{k\sigma} \bar{c}_{k\sigma}(i\omega_n)\, [-i\omega_n+\epsilon_k]\, c_{k\sigma}(i\omega_n)
             \\
         &+& {\cal S}_{\rm int}(\bar{c},c) + {\cal S}_{\rm loc}(f_\sigma)
           + \int_0^\beta \!\! d\tau \sum_{\{i\}\sigma} (V_i \bar{f}_{\sigma} c_{i\sigma} + c.c.)
\nonumber
\end{eqnarray}
where $\beta=1/T$ is the inverse temperature,
$c$ are the conduction electrons with dispersion $\epsilon_k$ on a regular lattice
with $\cal{N}$ sites,
${\cal S}_{\rm int}$ are the interactions within the conduction band, and
${\cal S}_{\rm loc}$ describes the $f$ electron impurity orbital with on-site energy
and repulsion. The sum $\sum_{\{i\}}$ runs over a set of lattice sites in the vicinity
of the impurity, and $V_i$ is the hybridization matrix element;
for a point-like impurity only a single $V_i$ is non-zero.
We can define a linear combination $c_0$ of conduction electron operators
to which the impurity couples:
$V c_{0\sigma} = \sum_{\{i\}} V_i c_{i\sigma}$ with $[c_0,c_0^\dagger]_+=1$.
After decoupling of ${\cal S}_{\rm int}$ and a suitable saddle-point approximation
of BCS type, all conduction electrons except $c_0$ can be integrated out.
(After the BCS approximation, the remaining integral is Gaussian and
can be performed exactly.)
To simplify notation we will restrict ourselves to BCS singlet pairing;
the arguments apply similarly to unitary triplet states, whereas
non-unitary triplet states require an additional coupling between the impurity
and the condensate spin moment.
Introducing a Nambu spinor $\Psi_0 =
(c_{0\uparrow},c_{0\downarrow}^\dagger)$
we obtain an action of the form:
\begin{eqnarray}
{\cal S} &=& - \frac{1}{\beta}
           \sum_{\omega_n} \bar{\Psi}_0(i\omega_n) \, G_0^{-1}(i\omega_n) \, \Psi_{0}(i\omega_n)  \nonumber\\
           &+& {\cal S}_{\rm loc}(f_\sigma)
           + \int d\tau (V \bar{f}_{\sigma} c_{0\sigma} + c.c.)
\label{impact}
\end{eqnarray}
Here, $G_0(i\omega_n)$ is the local conduction electron Green's function
at the impurity location.
In principle, a treatment of ${\cal S}_{\rm int}$ beyond mean-field
also generates a retarded self-interation for the $c_0$.
As we are mainly interested in a BCS-type host we will neglect this here.
Then, the properties of the bath are completely contained in $G_0$.
Explicitly we have
\begin{eqnarray}
G_0(i\omega_n) = \sum_k
\left(
\begin{array}{ll}
|h_k|^{-2}(i\omega_n \!-\!\epsilon_k) & h_k^{-2}\Delta_k \\
{h_k^\ast}^{-2} \Delta_k^\ast & |h_k|^{-2} (i\omega_n\!+\!\epsilon_k) \\
\end{array}
\right)^{-1} \!\!
\end{eqnarray}
where $\Delta_k$ is the complex gap function, and the function
$h_k$ contains the geometry of the impurity coupling to the host:
\begin{equation}
h_k = \sum_{\{i\}} e^{ikR_i} V_i / V \,.
\label{hk}
\end{equation}

\begin{figure}[t]
\epsfxsize=3.5in
\centerline{\epsffile{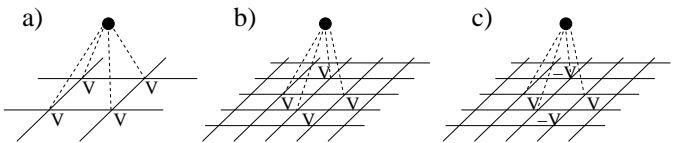}}
\caption{
Spatially extended impurities for which the anomalous
propagator in $G_0$ drops out, i.e.,
$\sum_k h_k^2 \Delta_k / D_k = 0$, for a square-lattice
$d$-wave superconductor.
The dot represents the impurity orbital,
the dashed lines show the non-zero hybridization paths (in the
notation of an Anderson model).
The situations b) and c) correspond to the
s-wave and d-wave linear combinations, i.e., screening channels,
of the four host sites, which occur in a {\em Kondo} model for a spatially
extended impurity, as discussed in Ref.~\protect\onlinecite{MVRB}.
}
\label{fig:imp}
\end{figure}

For a point-like impurity $h_k=1$, and the anomalous (off-diagonal) part of
$G_0$ is given by $G_0^{\rm a}(i\omega_n) = \sum_k \Delta_k/D_k$
(with $D_k = \omega_n^2 + \epsilon_k^2 + |\Delta_k|^2$),
which vanishes for unconventional superconductors with inversion symmetry
and Cooper pair angular momentum $l>0$, i.e., $p$-wave, $d$-wave or
higher symmetries.
(This also applies to $p_x+ip_y$ or $d_{x^2-y^2}+id_{xy}$ pairing states, but
not necessarily to $d+is$ states.)
Then, only the diagonal part of $G_0$ enters the impurity action (\ref{impact}),
which is completely determined by the local DOS.
For a spatially extended impurity hybridized with more than one site
the anomalous piece of $G_0$ is given by $\sum_k h_k^2 \Delta_k / D_k$
which still vanishes for some important situations:
Consider a $d$-wave superconductor, $\Delta_k = \Delta_0 (\cos k_x - \cos k_y)$,
and an impurity hybridized e.g. with four sites as in Figs.~1a,b,c.
In all cases the average over $h_k^2 \Delta_k/D_k$ vanishes for symmetry reasons.

At this point a brief comment on the experimental situation is in order.
Point-like impurities can be realized in layered superconductors with
out-of-plane impurity moments coupling to a single conduction electron orbital
only.
In contrast, in-plane impurities will typically couple to a number of sites, see Fig.~1.
The signs of the various hybridization matrix elements depend on the involved
orbitals of both impurity and host atoms.
We note that the situations in Figs.~1b,c were used in Refs. \onlinecite{tolya,MVRB}
to model the magnetic moment induced by a Zn impurity in a high-temperature
superconductor.
(There, a Kondo model for a spatially extended impurity was employed, leading to multiple
screening channels. In addition, the situation for Zn is complicated by the fact that
Zn, having a filled d shell, acts as a vacancy, but induces a magnetic moment in its
vicinity -- for details see Ref.~\onlinecite{MVRB}.)

Returning to the models discussion --
what about the behavior of a spatially extended Anderson impurity, where
$G_0^{\rm a}$ does {\em not} vanish?
Formally, we still have a single-channel model (\ref{impact}),
but now in the presence of both
a normal and an anomalous bath -- this is similar to a point-like impurity in
a $s$-wave superconductor.
As explained in Refs.~\onlinecite{sakaisc,fritz} the main effect of the anomalous
bath can be understood as a transverse charge pseudospin field which induces
an additional particle-hole asymmetry.

So far we have discussed Anderson impurity models.
For a single-site impurity the above discussion
applies identically to a Kondo model (via Schrieffer-Wolff transformation).
Multichannel physics arises only in a spatially extended Kondo impurity model
(independent of anomalous propagators),
see Refs.~\onlinecite{MVRB,ColemanMC}.

\begin{figure}[t]
\epsfxsize=3.1in
\centerline{\epsffile{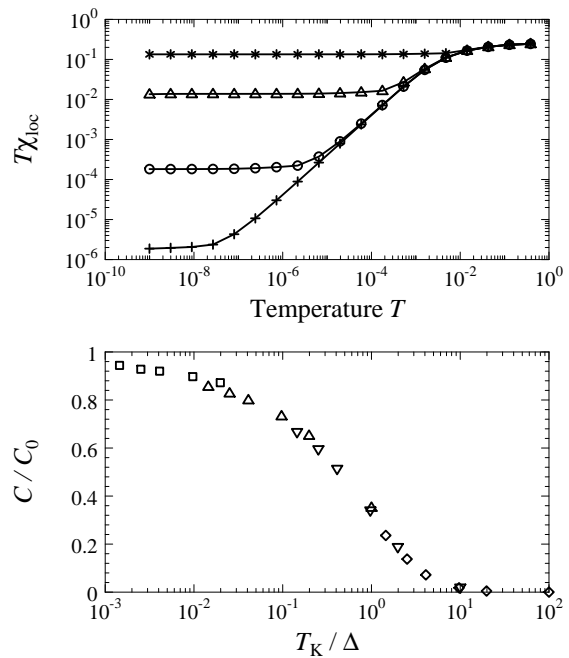}}
\caption{
Numerical results for the local susceptibility, obtained from a single-band
NRG for a (semiconducting) bath with the hard-gap DOS of a $d+id$ (or $p+ip$)
superconductor.
The host bandwidth is unity,
the NRG parameters are $\Lambda=3$ and $N_s=1000$.
a) $T\chi_{\rm loc}$ for different gap values
$\Delta = 10^{-2}, 10^{-3}, 10^{-4}, 10^{-5}$ (from top to bottom)
where the Kondo temperature $T_K$ of the model at $\Delta = 0$ is $3 \times 10^{-3}$.
b) The value of the local Curie constant $C$, normalized to its free-spin value 1/4,
as a function of the ratio $T_K / \Delta$.
Different symbols correspond to different values of $\Delta$.
(Deviations from a single universal curve are primarily due to NRG discretization
effects.)
The results can be compared with Fig. 4 of Ref.~\onlinecite{koga2}.
}
\label{fig:tchi}
\end{figure}

Now we comment on recent papers by Matsumoto and Koga \cite{koga1,koga2} about
a non-trivial ``orbital effect of Cooper pairs'' on the dynamics of point-like
impurities in $p+ip$ and $d+id$ superconductors.
They argue that, although the impurity only couples to the $s$-wave channel
of the conduction band, the $l\neq 0$ Cooper pairs mediate an indirect coupling
to higher angular momentum channels, leading eventually to a multichannel
Kondo problem.
For the cases of $p+ip$ and $d+id$ pairing, where the problem reduces to
two angular-momentum channels, they provided a numerical solution utilizing
a two-band numerical renormalization group (NRG) method.
Their findings show the absence of screening
even for large Kondo coupling, in contrast to the $s$-wave situation.
We believe that their mapping and results are likely to be correct, but allow
for a much simpler interpretation. Specifically:
(i) Our above arguments, which are rigorous for a point-like impurity in a
$d$-wave or unitary $p$-wave BCS superconductor,
show that only the local DOS enters the impurity problem.
(ii) The two-band model of Ref.~\onlinecite{koga2} [their Eq. (2.21)] is in fact
equivalent to a one-band model of the form (\ref{impact})
(in that it yields the same partition function and the same observables
related to the impurity degrees of freedom).
A first indication is that the impurity term in Eq. (2.21) of Ref.~\onlinecite{koga2}
mixes the two bands,
i.e., contains channel-flip terms (in contrast to true two-band models).
Formally, a one-band model can be obtained by integrating out the $a_{k,l=1}$
band in Eq. (2.4) of Ref.~\onlinecite{koga2} exactly -- this will generate a {\em normal} self-energy
for the $a_{k,l=0}$ fermions and induce the quasiparticle gap \cite{wavef}.
(iii) The local DOS of a $p+ip$ or $d+id$ superconductor displays a hard gap,
thus we expect the qualitative impurity properties to be those of a hard-gap
system \cite{hardgap}.
In particular, for a p-h symmetric band there is no screening even at large
Kondo coupling, consistent with the results of Ref.~\onlinecite{hardgap}.
However, in the presence of p-h asymmetry there will be a first-order transition
to a screened phase as the Kondo coupling is increased \cite{hardgap}.
This transition was not found in Refs.~\onlinecite{koga1,koga2},
instead the authors claim that their findings persist in the presence of p-h asymmetry
(but numerical results are not shown).
We believe this claim to be incorrect.
(iv) We have performed {\em one-band} NRG calculations using the local quasiparticle DOS of a
$d+id$ (or $p+ip$) superconductor,
and have essentially reproduced the results for the local susceptibility
of Ref.~\onlinecite{koga2}, see Fig. 2
\footnote{Due to the different truncation schemes in
the one-band and two-band NRG codes {\em identical} results cannot be obtained.}.
Taken together, the physics of point-like impurities described in Refs.~\onlinecite{koga1,koga2}
is completely contained in a Kondo model
where the impurity is coupled to a single band with a (fully gapped) local DOS
of a non-superconducting host \cite{wavef},
and the ``orbital effect of the Cooper pairs'' does not exist.
(This is different for spatially extended Kondo impurities where true multiband
effects obtain \cite{koga3}.)
The discussion in Sec.~III~B of Ref.~\onlinecite{koga2} is formally correct, but
misses that the mechanism leading to the doublet ground state of a strongly coupled
Kondo impurity is identical for any p-h symmetric hard-gap system \cite{hardgap}.

In summary, we have argued that in unconventional BCS superconductors
the properties of a Kondo impurity are not influenced by the presence
of anomalous propagators,
provided that $\sum_k h_k^2 \Delta_k / D_k = 0$ (\ref{hk}) --
this is e.g. the case for point-like impurities and a vanishing local superconducting
order parameter.
Under these conditions, using only the local density of states as
bath input quantity for impurity calculations, as done in Refs.~\onlinecite{tolya,MVRB},
is exact.



We thank R. Bulla, S. Florens, M. Kir\'{c}an, and A. Schiller for helpful discussions.
This research was supported by the Deutsche Forschungsgemeinschaft through
the Center for Functional Nano\-structures Karlsruhe.



\begin{thebibliography}{}

\bibitem{sakaisc} K.~Satori, H.~Shiba, O.~Sakai, and Y.~Shimizu,
J. Phys. Soc. Jpn. {\bf 61}, 3239 (1992).

\bibitem{withoff}
D.~Withoff and E.~Fradkin, Phys. Rev. Lett. {\bf 64}, 1835 (1990).

\bibitem{cassa}
C.~R.~Cassanello and E.~Fradkin,
Phys. Rev. B {\bf 53}, 15079 (1996) and {\bf 56}, 11246 (1997).


\bibitem{tolya} A.~Polkovnikov, S. Sachdev, and M. Vojta, \prl {\bf 86}, 296 (2001).

\bibitem{MVRB} M. Vojta and R. Bulla, \prb {\bf 65}, 014511 (2002).

\bibitem{MVMK} M. Vojta and M. Kir\'{c}an, \prl {\bf 90}, 157203 (2003).

\bibitem{koga1} M. Matsumoto and M. Koga, J. Phys. Soc. Jpn. {\bf 70}, 2860 (2001).  
\bibitem{koga2} M. Matsumoto and M. Koga, \prb {\bf 65}, 024508 (2002). 
\bibitem{koga3} M. Koga and M. Matsumoto, J. Phys. Soc. Jpn. {\bf 71}, 943 (2002) and
                                          \prb {\bf 65}, 094434 (2002). 

\bibitem{wavef}
As our arguments are based on (exactly) integrating out certain conduction electrons,
certain observables involving the conduction band cannot be obtained directly from
Eq.~\protect\ref{impact}.
From the solution of the local impurity problem \protect\ref{impact} one obtains the
T matrix, which then allows to directly calculate all conduction electron observables.

\bibitem{hardgap}
K. Chen and C. Jayaprakash, Phys. Rev. B {\bf 57}, 5225 (1998).

\bibitem{fritz}
L. Fritz and M. Vojta, Phys. Rev. B {\bf 70}, 214427 (2004).

\bibitem{ColemanMC} P.~Coleman and A.~M.~Tsvelik, \prb {\bf 57}, 12757 (1998).

\end{thebibliography}
\end{document}